\documentclass[aps,prl,floatfix,epsfig,twocolumn,showpacs,preprintnumbers]{revtex4}
\usepackage{amssymb}
\usepackage{graphicx}
\usepackage{amsmath}

\begin{document}

\title{Ground State Properties of Simple Elements from GW Calculations}
\author{Andrey Kutepov$^{\ast }$, Sergey Y. Savrasov$^{\ast }$, Gabriel
Kotliar$^{\dag }$}
\affiliation{$^{\ast }$Department of Physics, University of California, Davis, CA 95616}
\affiliation{$^{\dag }$Department of Physics, Rutgers University, Piscataway, NJ 08856}

\begin{abstract}
A novel self--consistent implementation of Hedin's GW perturbation theory is
introduced. This finite--temperature method uses Hartree--Fock wave
functions to represent Green's function. GW equations are solved with full
potential linear augmented plane wave (FLAPW) method at each iteration of a
self--consistent cycle. With our approach we are able to calculate total
energy as a function of the lattice parameter. Ground state properties
calculated for Na, Al, and Si compare well with experimental data.
\end{abstract}

\maketitle

Density--functional theory (DFT)\cite{hk65} in its local density
approximation (LDA) or generalized gradient approximation (GGA) is a
widely used method to calculate ground state properties of solids.
But this theory is not always good in situations where electronic
correlations are essential. The problem is then arising that
existing variants of DFT cannot be improved systematically (or at
least it is very difficult to accomplish that). It is therefore very
desirable to have a method capable to deal with ground state
properties and at the same time allowing their systematical
improvement. This flexibility is obviously present in methods based
on diagrammatic expansion and one of them, Hedin's GW approach, is
becoming computationally accessible during last years. However,
while the GW method has become standard for studying excitation
spectra, the possibility of using it for calculating total energies
and other ground state properties is not well established. This
question has been addressed only in a few works. Holm and von Barth
\cite{prb_57_2108} have performed self--consistent GW calculations
for the homogeneous electron gas (HEG) and concluded that in spite
of a bad electronic structure obtained in their self--consistent
(SC) calculations the total energy was quite close to the result of
a quantum Monte Carlo \cite{prl_45_566} study. Garcia--Gonzales and
Godby \cite{prb_63_075112} also have obtained good total energies of
HEG (including spin--polarized case) in their SC GW calculations.
Stan, Dahlen and van Leeuwen \cite{epl_76_298} have applied SC GW to
calculate total energies of atoms and molecules. They conclude that
GW calculations should be done self-consistently in order to obtain
physically meaningful and unambiguous energy differences. Miyake,
Aryasetiawan, Kino, and Terakura \cite{prb_64_233109} have studied
ground state properties of sodium and aluminum using
Galitskii--Migdal formula \cite{jetp_139_96} with a model spectral
function. Their equilibrium volumes and bulk modulus appeared to be
slightly overestimated as compared to the experiment, but some
improvement over LDA results was reported. Miyake \textit{et al.}
\cite{prb_66_245103} have applied the total energy formula due to
Luttinger and Ward \cite{pr_118_1417} to the calculation of
equilibrium lattice constants in Na and Si. Their one--shot type
results appeared to be very close to the experimental data. However,
to our knowledge fully self--consistent GW calculations of the total
energy and the ground state properties for real solids have not yet
been carried out.

To address this question we have implemented a variant of GW method which
allows us to calculate total energies. The key ingredients of our
implementation are the following. First, we use full potential linear
augmented plane waves method (FLAPW)\cite{jcm03} to find the solutions of
the Hartree--Fock (HF) equations which serve as a basis for the expansion of
one--electron Green's function. Second, we have found it vital to use
Mazubara's time $\tau $--mesh to calculate correlated part of the
self--energy. It appears that it is very difficult to obtain comparable
accuracy in total energy using self--energies calculated in frequency
domain. Third, our calculations are self--consistent.

In our imlementation there are two formulas for the total energy. They
differ in how to find its exchange--correlation part. The first
implementation uses the Galitskii-Migdal formula, i.e. convolution of the
Green function $\mathcal{G}_{\lambda ^{\prime }\lambda }^{\alpha }(\mathbf{k}%
;\tau )$ and the self--energy $\Sigma _{\lambda \lambda ^{\prime }}^{\alpha
}(\mathbf{k};-\tau )$

\begin{equation}
E_{xc}=\frac{1}{2}\sum_{\alpha \mathbf{k}}\sum_{\lambda \lambda ^{\prime
}}\int \Sigma _{\lambda \lambda ^{\prime }}^{\alpha }(\mathbf{k};-\tau )%
\mathcal{G}_{\lambda ^{\prime }\lambda }^{\alpha }(\mathbf{k};\tau )d\tau ,
\label{E_c}
\end{equation}%
where $\alpha $ is a spin index; $\mathbf{k}$ denotes points in the
Brillouin zone, $\lambda ,\lambda ^{\prime }$ are band indexes. Second
formula follows from the equation of motion for the one--particle Green
function:

\begin{align}
E_{xc}& =-\frac{1}{2}E_{kin}+\frac{1}{2}\sum_{\alpha \mathbf{k}%
}\sum_{\lambda \lambda ^{\prime }}\mathcal{G}_{\lambda \lambda ^{\prime
}}^{\alpha }(\mathbf{k};\beta )V_{\lambda ^{\prime }\lambda }^{H,\alpha }(%
\mathbf{k})+\frac{1}{2}\mu Q  \notag  \label{E_xc3} \\
& +\frac{1}{2}\sum_{\alpha \mathbf{k}}\sum_{\lambda }\frac{\partial \mathcal{%
G}_{\lambda \lambda }^{\alpha }(\mathbf{k};\tau )}{\partial \tau }|_{\tau
=\beta },
\end{align}%
where $E_{kin}$ is a kinetic energy of the electrons; $V_{\lambda ^{\prime
}\lambda }^{H,\alpha }(\mathbf{k})$ denote matrix elements of the Hartree
potential; $\mu $ is a chemical potential; $\beta =1/T$, and $Q$ is the
number of electrons.

We check the convergence of the total energy with respect to the number of
points on $\tau $--mesh by comparing the result from Eq. (\ref{E_xc3}) with
the result from Eq. (\ref{E_c}). We find that the formula (\ref{E_c}) gives
much more stable results than Eq.(\ref{E_xc3}).

\begin{figure}[t!]
\centering \rotatebox{-90}{
\includegraphics[width=6.0 cm]{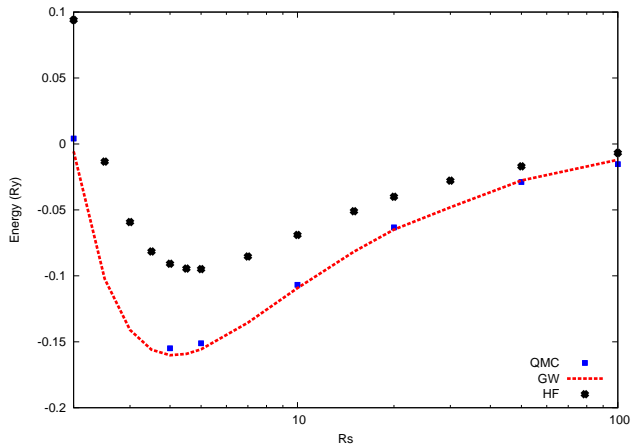} }
\caption{Total energy of HEG with respect to $r_{s}$ obtained in HF and GW
approximations. Comparison is made with QMC results.\protect\cite{prl_45_566}
}
\label{heg_etot}
\end{figure}

As a test of our newly developed code we have applied it to calculate the
total energy for the homogeneous electron gas. As we have already mentioned,
this was already done previously (\cite{prb_57_2108}, \cite{prb_63_075112}),
but in those calculations zero temperature formalism was utilized. In Fig. %
\ref{heg_etot} the result from our finite temperature (300K)
approach is shown in comparison with the accurate Quantum Monte
Carlo data obtained by Ceperly and Alder.\cite{prl_45_566} We have
also calculated total energy of HEG using the HF approach. As it is
seen, SCF GW total energies are in very good agreement with QMC
simulations and show great improvement compared with the HF results.

All our calculations for real solids have been performed for temperature
T=2000K. For our LDA and GGA studies we used the exchange--correlation
functionals from Refs. [\onlinecite{prb_45_13244}], [\onlinecite{pcvjpsf92}]
respectively. We have listed most important calculational parameters for all
studied systems in Table \ref{par_calc}.
\begin{table}[t]
\caption{Parameters of calculations}
\label{par_calc}
\begin{center}
\begin{ruledtabular}
\item[]\begin{tabular}{@{}c c c c}  & Na &Al & Si\\
\hline
$\mathbf{k}$-mesh  &$8 \times 8 \times 8$  &$8 \times 8 \times 8$  &$5 \times 5 \times 5$\\
$N_{bands}$&23  &15 &40\\
$N_{\tau}$   &80 &48  & 40  \\
$N_{LAPW}$& 23  & 38&73 \\
$N_{PB}$  &135  & 190  &513 \\
Semicore  & 2s2p  &  2s2p& -  \\
\end{tabular}
\end{ruledtabular}
\end{center}
\end{table}

\begin{figure}[b]
\centering
\includegraphics[width=10 cm]{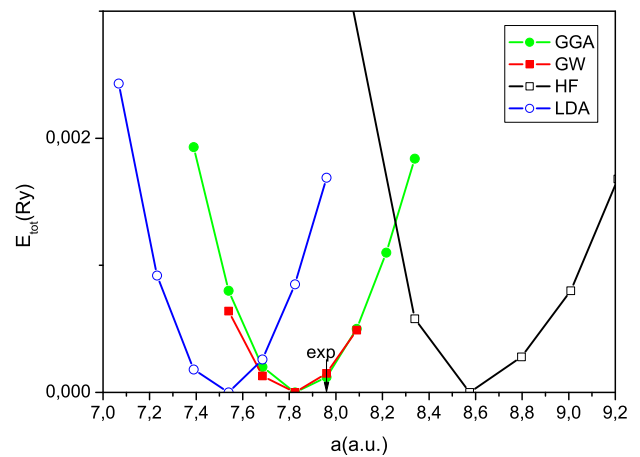}
\caption{Total energy of Na as a function of lattice parameter. Arrow
indicates the experimental lattice parameter.\protect\cite{prl_60_1558}}
\label{Na_etot}
\end{figure}
They include the k--point sampling in the Brillouin zone, the size
of LAPW basis, $N_{LAPW}$, the size of product basis, $N_{PB}$, the
number of bands used for the representation of correlated part of
the Green function, and the number of points on Mazubara's time
mesh. Absolute convergence of the total energy in our calculation is
a bit worse than the one typical for LDA calculation (1mRy/atom),
but it should also be noted that the convergence of DOS is reached
faster than the convergence of the total energy. Furthermore, there
is some dependence of our predicted equilibrium volumes with respect
to some specific parameters. The most critical one is the number of
k--points. The influence of other parameters mostly results in a
rigid shift of the total energy with no much change in the position
of a minimum.

As an example, Fig. \ref{Na_etot} presents our calculated total energy
versus lattice parameter for Na while Table \ref{GRP} contains the
calculated ground state properties for all studied elements: Na, Al, and Si.
Here we compare the results of the LDA, GGA, HF, and GW approximations. All
these calculations are self--consistent. As it is seen there is the same
tendency for all materials with respect to the predicted lattice parameters:
the LDA gives too small values, while the HF approach predicts them to be
too big (especially for metals). The GGA result is better, though for Al
this approach leads to a bit expanded volume. We should stress that the GW
result is very accurate (especially for aluminum and silicon). For sodium,
both the GW and the GGA results are quite similar. We should also note that
for a semiconductor (Si) the HF approach is quite competitive in accuracy
with the GW but it gives a little expanded volume which is not good if one
thinks of additional expansion connected to the lattice vibrations.

Bulk moduli do not show such a consistent improvement when calculated within
the GW approach. Only for silicon we have found a remarkable agreement with
experiment. We think that the reason is that our results are not well
convergent with respect to the $\mathbf{k}$--points sampling used but it is
too expensive computationally to study its full convergence.

\begin{table}[t]
\caption{Equilibrium lattice parameter $a_{0}$(a.u.) and bulk modulus $B_{0}$%
(GPa) of Al, Na, and Si compared to experimental data \protect\cite%
{prb_37_790,prb_31_5327}}
\label{GRP}
\begin{center}
\begin{ruledtabular}
\item[]\begin{tabular}{@{}c c c c c c c}  & \multicolumn{2}{c}{Na} & \multicolumn{2}{c}{Al}&\multicolumn{2}{c}{Si} \\
\hline
   & $a_{0}$  & $B_{0}$  &$a_{0}$   &$B_{0}$  & $a_{0}$  & $B_{0}$   \\
$LDA$  & 7.53  & 87  &7.585  &82.5  & 10.01 & 91.2   \\
$GGA$  & 7.83  & 71.5  &7.72  &73.3  &10.14  & 93.5  \\
$HF$  & 8.62  & 50  & 7.82 & 82.5 &10.34  & 97.0   \\
$GW$  &7.82   &63.0   & 7.64 &86.5  &10.17  &100.7   \\
$Exp$  & 7.96  &68.1   &7.65  &72.16  &10.26  & 99  \\

\end{tabular}
\end{ruledtabular}
\end{center}
\end{table}

%\begin{figure}[t!]
%\centering
%\includegraphics[width=10 cm]{Na_DOS}
%\caption{DOS of Na calculated in GGA, HF and GW approximations.
%Chemical potential is placed at zero energy. The arrow shows the
%experimental position of the valence band bottom.}\label{Na_scf_100}
%\end{figure}

\begin{figure}[b!]
\centering
\includegraphics[width=10 cm]{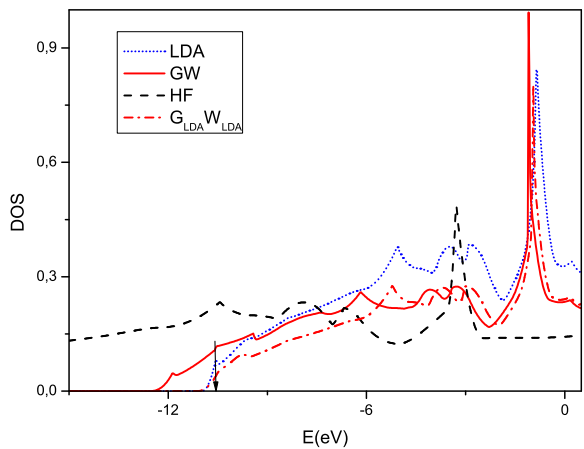}
\caption{DOS of Al calculated in GGA, HF and GW approximations. Chemical
potential is placed at zero energy. The arrow shows the experimental
position of the valence band bottom.}
\label{Al_scf_100}
\end{figure}

As a byproduct of our study we have investigated the influence of
the self--consistency on the excitation spectra. Despite it was
already noticed long ago \cite{prb_34_5390} that the GW
approximation can be quite useful for calculating one--electron
spectra, there is still some ambiguity with respect to the effect of
self consistency. Usually, GW calculations are exceedingly demanding
and only one--shot variant is used to calculate band widths or band
gaps. Calculated on top of LDA or GGA such one--shot quasiparticle
band structures appeared to be in much better agreement with
experimental data when compared to the LDA ones
\cite{prb_57_2108,prl_81_1662,rpp_61_237,prb_74_045102,prb_76_165106}.
Later it was shown that self consistency within GW gives too large
band gaps for semiconductors and insulators
\cite{prl_81_1662,prb_74_045102,prl_99_246403}. Shishkin, Marsman
and Kresse \cite{prl_99_246403} have argued recently that the
inclusion of vertex corrections is required to obtain accurate band
gaps in the framework of SC GW. However this question of
self--consistency still remains as a few other works aimed at
self--consistent GW\ calculations used some simplifications.
Sch\"{o}ne and Eguiluz \cite{prl_81_1662} utilized pseudopotential
approach. Ku and Equiluz \cite{prl_89_126401} have applied
all--electron method but with 'diagonal' approximation to Green's
function. The calculations by Zein, Savrasov, and Kotliar
\cite{prl_96_226403} were performed with atomic sphere approximation
(ASA). Bruneval, Vast, and Reining \cite{prb_74_045102} have
combined the self--consistent screened exchange plus Coulomb hole
(COHSEX) with one--shot GW. Shishkin and Kresse \cite{prb_75_235102}
have updated self--consistently only the eigenvalues in the Green's
function. Kotani, van Schilfgaarde, and Faleev \cite{prb_76_165106}
applied a strategy of finding an effective Hamiltonian so that
further perturbative contributions would be as small as possible.
Their approach has some justifications though it is not true answer
for the self--consistent GW.

%\begingroup
%\squeezetable
\begin{table*}[tbp]
\caption{Band gap (energies in eV) for Si. The values in round
brackets were obtained with SC in eigen values only. Experimental
band gap is 1.17ev.} \label{si_gap}
\begin{center}
\begin{ruledtabular}
\item[]\begin{tabular}{@{}c c c c c c c c}  & Ref.[\onlinecite{prl_81_1662}] &Ref.[\onlinecite{prl_89_126401}] &
 Ref.[\onlinecite{prl_96_226403}] &Ref.[\onlinecite{prb_76_165106}] & Ref.[\onlinecite{prb_74_045102}]&
 Ref.[\onlinecite{prl_99_246403}] & Present work \\
\hline

$LDA$  & 0.53  & 0.52  &  &0.46  & 0.51 &   & 0.46  \\
$GGA$  &   &   &  &  &  &   & 0.54  \\
$HF$  &   &   &  &  &  &   & 6.27  \\
$G_{LDA}W_{LDA}$  &1.34   & 0.85  & 0.86  &0.98  & 1.14  &  & 0.86 \\
$G_{HF}W_{HF}$  &   &   &  &   &   &  &2.69  \\
$G_{COHSEX}W_{COHSEX}$ &   &   &  &  & 1.56  &  &  \\
$GW_{LDA}$ &   &   &  &  &  &1.28(1.20)  &  \\
$QPscGW$ &   &   &  &1.25(1.14)  & 1.47 & 1.41  &  \\
%$QPscGW+e-h$  &   &   &  &  &  &1.24   &   \\
$GW$  &1.91   &1.03   & 1.10 &  &  &   & 1.55  \\

\end{tabular}
\end{ruledtabular}
\end{center}
\end{table*}
%\endgroup

To address the issue of an influence of the self--consistency on the
calculated electronic structure we perform SC GW calculations and
compare the results with the results from the one--shot GW
calculations based on DFT self--consistent calculations. As example
of calculated DOS for metals, we present our calculated electronic
structure of Al in Fig.\ref{Al_scf_100}. We are especially concerned
with the valence band width for this material. First observation
(which may be trivial one) is that HF approximation gives too big
band width. Second observation is also quite known for simple
metals: both LDA and GGA calculations give slightly expanded band
width as compared to experimental data. As it is seen, our non SC GW
band width for Al is much closer to the LDA result and to
experimental data than SC GW result which is about 10\% too big as
compared to the experimental works by Lyo and
Plummer\cite{prl_60_1558} and by Livins and Schnatterly
\cite{prb_37_6731}. Since Aluminum is a free--electron metal, we
think that the above result is consistent with the results by Holm
and Barth \cite{prb_57_2108}, and by Shirley\cite{prb_54_7758}
obtained for the homogeneous electron gas, where the calculated by
SC\ GW band width was found to be too big as compared to the non
self--consistent one. Based on the work by Shirley we expect that
higher order vertex corrections will bring the calculated band width
in closer agreement with experiment.

In Table \ref{si_gap} we have collected the calculated fundamental
gaps for silicon obtained in earlier works along with our results.
Our band gap from non SC GW calculation is quite close to the
results of others. However our band gap from SC GW appears to be a
bit wider than it is usually obtained. We would stress however that
neither of previous calculations are approximation free and that
partially SC results obtained in the works \cite{prb_76_165106,
prb_74_045102, prl_99_246403} also show a trend in increasing the
band gap. In this respect it would be very interesting to know what
effect will bring the vertex correction on the band gap in Si. This
question we hope to answer in the nearest future.

In summary, we have presented a self--consistent realization of the GW
method and its performance for the evaluation of the total energy and
ground--state properties. For the materials we studied (Na, Al, and Si) the
GW approach leads to consistently better description of equilibrium volume
than LDA, GGA or HF approaches do. Our results for the electronic structure
of the above mentioned materials are in qualitative agreement with earlier
works, but in general we should conclude that some deterioration in
calculated band widths and gaps is seen when we are trying to use
all--electron, full--potential method in connection with "classical"
self--consistent GW approach. In this respect our conclusion is the same as
one by Holm and Barth \cite{prb_57_2108}: self--consistent GW approach can
produce good total energies but it is not so accurate for the one--electron
spectra.

%\bibliographystyle{plain}
%\bibliography{Method,Actinides}

\end{document}